\documentclass[%
superscriptaddress, groupedaddress, unsortedaddress, runinaddress,
frontmatterverbose, preprint, showpacs,preprintnumbers, nofootinbib,
nobibnotes, bibnotes, amsmath,amssymb, aps, prd,
floatfix, ]{revtex4-1}

\usepackage{graphicx}% Include figure files
\usepackage{dcolumn}% Align table columns on decimal point
\usepackage{bm}% bold math
\usepackage[mathlines]{lineno}% Enable numbering of text and display math
\begin{document}

\preprint{}

\title{Fluctuations and correlations of hot QCD matter in an external magnetic field}% Force line breaks with \\

\author{Wei-jie Fu}
 \email{fuw@brandonu.ca}
\affiliation{%
 Department of Physics, Brandon University, Brandon,
Manitoba, R7A 6A9 Canada
}%

\date{\today}% It is always \today, today,

\begin{abstract}
Effects of an external magnetic field on the fluctuations of quark
number, fluctuations and correlations of conserved charges,
including baryon number, electric charge and strangeness, are
studied in the 2+1 flavor Polyakov--Nambu--Jona-Lasinio model. We
find that magnetic field increases fluctuations and correlations in
the regime of chiral crossover. It makes the transition of quadratic
fluctuations more abrupt, and the peak structure of quartic
fluctuations more pronounced. Our calculations indicate that
$\chi_{4}^{B}/\chi_{2}^{B}$, $\chi_{4}^{Q}/\chi_{2}^{Q}$, and
$\chi_{11}^{BQ}$ are very sensitive to the external magnetic field
and maybe can be used for probes for the strong magnetic field
produced in the early stage of noncentral collisions.
\end{abstract}


%Use showkeys class option if keyword
\maketitle

%\tableofcontents

\section{Introduction}
\vspace{5pt}

Studies of influences of strong magnetic fields on hot QCD matter
have attracted lots of attentions in recent years. It is estimated
that the strength of magnetic fields produced in noncentral
relativistic heavy ion collisions can reach up to $eB\sim 0.1
m_{\pi}^{2}$ for SPS, $eB\sim m_{\pi}^{2}$ for RHIC, and $eB\sim 15
m_{\pi}^{2}$ for LHC~\cite{Skokov2009}. Furthermore, The strength of
magnetic fields produced in the early universe where the electroweak
phase transition happened, may even reach up to $eB\sim 200
m_{\pi}^{2}$~\cite{Vachaspati1991}. An interesting phenomenon
related to magnetic fields in noncentral heavy ion collisions is the
electric charge separation along the system's orbital angular
momentum axis~\cite{Abelev2009a,Abelev2009b}. The observed electric
charge separation can be explained as the chiral magnetic effect,
where an electric current is induced along the direction of magnetic
field and positive charges are separated from negative ones in
parity-odd
domains~\cite{Fukushima2008,Kharzeev2009,Kharzeev2010,Fukushima2009,Fu2011b,Fu2011}.

Effects of strong magnetic fields on the chiral and deconfinement
phase transitions have been studied within the frameworks of
effective models and lattice-QCD simulations. In a two-flavor
Polyakov--Nambu--Jona-Lasinio (PNJL) model, it was observed that the
external magnetic field works as a catalyzer of dynamical symmetry
breaking and the critical temperature increases with the strength of
$B$~\cite{Fukushima2010}. Calculations of a linear sigma model
coupled to quarks and to the Polyakov loop indicated that the chiral
and deconfinement lines split and both chiral and deconfining
critical temperature increase with $B$~\cite{Mizher2010}.
Furthermore, it was also found within effective models that the
transition strength increases when the magnetic field strength is
increased~\cite{Skokov2012}. Based on a renormalization group (RG)
analysis, Fukushima and Pawlowski found a reason for the magnetic
catalysis for the dynamical chiral symmetry
breaking~\cite{Fukushima2012}, which was observed in various
effective model calculations: when a strong magnetic field is
applied, the transverse dynamics of charged particles is frozen and
the dimensional reduction takes place. In this case the RG flow of
the dimensionless coupling results in a divergent coupling, no
matter how small the initial coupling is. However, this happens only
when the initial coupling exceeds a critical value in the case
$B=0$.

Two flavor lattice-QCD simulations in Ref.~\cite{DElia2010} found
that both the deconfinement and chiral restoring critical
temperatures increases with $B$, but they did not find the splitting
of the chiral phase transition. They also found that the transition
strength increases with increasing $B$. However, another
state-of-the-art lattice simulations with 2+1 flavors of stout
smeared staggered quarks with physical masses found the transition
temperature decreases with increasing magnetic field, which is
inconsistent with various model calculations, but they also found
the transition strength increasing mildly with $B$~\cite{Bali2012}.
With the framework of QCD effective potential for the homogeneous
Abelian gluon field, it was found that the strong magnetic field
catalyze the deconfinement transition~\cite{Galilo2011}.

Within the framework of 2+1 flavor Polyakov-loop improved NJL
model~\cite{Fu2008}, in this work we study the effects of the
magnetic field on the fluctuations and correlations of quark number
and conserved charges, e.g., baryon number, electric charge, and
strangeness. Fluctuations and correlations of conserved charges are
sensitive to the degrees of freedom of the thermal strongly
interacting matter and behave quite differently between the hadronic
and quark gluon plasma (QGP) phases~\cite{Jeon2000,Koch2005}.
Fluctuations and correlations are usually enhanced near the QCD
phase transitions, and are related to the critical behavior of the
QCD
thermodynamics~\cite{Stephanov1998,Hatta2003,Jeon2004,Stephanov2004,Skokov2010}.
Furthermore, fluctuations and correlations of conserved charges can
be measured with event-by-event fluctuations in heavy ion collision
experiments~\cite{Koch2005,Jeon2004,Abelev2009,Aggarwal2010}, and so
they are valuable probes of the deconfinement and chiral restoring
phase
transitions.~\cite{Ejiri2006a,Ejiri2006b,Karsch2006,Stokic2009}.

The fluctuations of conserved charges as well as the correlations
among them without external magnetic field have been studied in the
2+1 flavor PNJL model~\cite{Fu2010a,Fu2010b}. The calculated results
were compared with lattice simulations performed with an improved
staggered fermion action with almost physical up and down quark
masses and a physical value for the strange quark mass. It was found
that the calculated results of effective model are well consistent
with those obtained in lattice simulations~\cite{Fu2010a}, which
indicates that the 2+1 flavor PNJL model is well suitable for the
calculations of the cumulants of conserved charge multiplicity
distributions. This computation of effective model was also extended
to study the fluctuations and correlations near the QCD critical
point and many interesting results are obtained~\cite{Fu2010b}.
Since the fluctuations and correlations of conserved charges can be
observed in heavy ion collision experiments, it is expected that
effects of the strong magnetic fields produced in the early
evolution stage of noncentral collisions are imprinted onto these
observables. In another word, maybe we can employ the fluctuations
and correlations to infer the presence or information of the
magnetic field. But before this idea comes true, we have to study
the influences of an external magnetic field on the fluctuations and
correlations of quark number and conserved charges, which is our
focus in this work.

The paper is organized as follows. In section \ref{model} we
introduce the effective model and the fluctuations and correlations
of conserved charges. In Sec. \ref{phasetransition} we show the
calculated results of the deconfinement and chiral restoring phase
transitions. In Sec. \ref{fluctuationsquark} we give our calculated
results of fluctuations of light quarks (up and down quarks) and
strange quarks in an external magnetic field. Section
\ref{conservedcharges} shows the fluctuations of conserved charges
and correlations among them in an external magnetic field. In Sec.
\ref{summary} we present our summary and conclusions.

\section{Effective model}
\label{model}

In this work, we employ the 2+1 flavor Polyakov-loop improved NJL
model to study the fluctuations and correlations of conserved
charges in an external magnetic field. We begin with the 2+1 flavor
PNJL model (for more details about the PNJL model, see
Ref.~\cite{Fu2008} and references therein), whose Lagrangian density
reads
\begin{eqnarray}
\mathcal{L}_{\mathrm{PNJL}}&=&\sum_{f=u,d,s}\bar{\psi_{f}}(i\gamma_{\mu}D^{\mu}_{f}+\gamma_{0}
 \mu_{f}-m_{0f})\psi_{f}
 +G\sum_{a=0}^{8}\Big[(\bar{\psi}\tau_{a}\psi)^{2}
 +(\bar{\psi}i\gamma_{5}\tau_{a}\psi)^{2}\Big] \nonumber \\
&&-K\Big[\textrm{det}_{f}\big(\bar{\psi}(1+\gamma_{5})\psi\big)
 +\textrm{det}_{f}\big(\bar{\psi}(1-\gamma_{5})\psi\big)\Big]
 -\mathcal{U}(\Phi,\Phi^{*} \, ,T),\label{lagragian}
\end{eqnarray}
where the covariant derivative $D_{\mu
f}=\partial_{\mu}+iq_{f}ea_{\mu}-iA_{\mu}$ couples the quark field
$\psi_{f}$ to the electromagnetic field (here we denote it as
$a_{\mu}$) and the background gluon field $A_{\mu}$. $q_{f}$
$(f=u,d,s)$ is the electric charge in unit of elementary electric
charge $e$ for the quark of flavor $f$. Usually in the Polyakov-loop
improved effective models, we only keep the temporal component of
the background gluon field, i.e., $A^{\mu}=\delta_{0}^{\mu}A^{0}$,
with
\begin{equation}
A^{0}=g\mathcal{A}^{0}_{a}\frac{\lambda_{a}}{2},
\end{equation}
where $\lambda_{a}$ are the Gell-Mann matrices in color space.
$m_{0f}$ and $\mu_{f}$ in Eq. (\ref{lagragian}) are the current
quark masses and the quark chemical potentials, respectively. We
choose $m_{0s}>m_{0u}=m_{0d}\equiv m_{0l}$ throughout this work,
which breaks the $SU(3)_f$ symmetry. In addition to the quark
chemical potentials, we will encounter chemical potentials for
conserved charges in the following discussions, e.g., $\mu_{B}$,
$\mu_{Q}$, and $\mu_{S}$, which are the chemical potentials for the
baryon number, electric charge, and strangeness, respectively. They
are related with the quark chemical potentials through the following
relations:
\begin{equation}
\mu_{u}=\frac{1}{3}\mu_{B}+\frac{2}{3}\mu_{Q},\quad
\mu_{d}=\frac{1}{3}\mu_{B}-\frac{1}{3}\mu_{Q},\quad\textrm{and}\quad
\mu_{s}=\frac{1}{3}\mu_{B}-\frac{1}{3}\mu_{Q}-\mu_{S}.
\end{equation}

$\mathcal{U}\left(\Phi,\Phi^{*},T\right)$ in Eq. (\ref{lagragian})
is the Polyakov-loop effective potential, which is usually a
function of an order parameter of the deconfinement phase
transition, the traced Polyakov loop
$\Phi=(\mathrm{Tr}_{c}L)/N_{c}$, and its conjugate
$\Phi^{*}=(\mathrm{Tr}_{c}L^{\dag})/N_{c}$. The Polyakov loop is
linked to the background gluon field through
\begin{equation}
L\left(\vec{x}\right)=\mathcal{P}\exp\left[i\int_{0}^{\beta}d\tau\,
A_{4}\left(\vec{x},\tau\right)\right] =\exp\left[i \beta A_{4}
\right]\, ,
\end{equation}
where $\mathcal{P}$ denotes path ordering; $\beta=1/T$ is the
inverse of temperature and $A_{4}=iA^{0}$. The functional form and
parametrization of the Polyakov-loop effective potential in the PNJL
model are determined phenomenologically by fitting the
thermodynamical behavior of the Polyakov dynamics for the pure gauge
field. In this work, we employ the Polyakov-loop effective potential
which is a polynomial in $\Phi$ and $\Phi^{*}$~\cite{Ratti2006a},
given as
\begin{equation}
\frac{\mathcal{U}\left(\Phi,\Phi^{*},T\right)}{T^{4}} =
-\frac{b_{2}(T)}{2}\Phi^{*}\Phi -\frac{b_{3}}{6}
(\Phi^{3}+{\Phi^{*}}^{3})+\frac{b_{4}}{4}(\Phi^{*}\Phi)^{2} \, ,
\end{equation}
with
\begin{equation}
b_{2}(T)=a_{0}+a_{1}\left(\frac{T_{0}}{T}\right)+a_{2}
{\left(\frac{T_{0}}{T}\right)}^{2}
+a_{3}{\left(\frac{T_{0}}{T}\right)}^{3}.
\end{equation}
Parameters in the effective potential are $a_{0}=6.75$,
$a_{1}=-1.95$, $a_{2}=2.625$, $a_{3}=-7.44$, $b_{3}=0.75$,
$b_{4}=7.5$, and $T_{0}=270\,\mathrm{MeV}$, which are fixed
according to lattice simulations. Furthermore, five parameters in
the quark sector of the model are determined to
$m_{0}^{l}=5.5\;\mathrm{MeV}$, $m_{0}^{s}=140.7\;\mathrm{MeV}$,
$G\Lambda^{2}=1.835$, $K\Lambda^{5}=12.36$, and
$\Lambda=602.3\;\mathrm{MeV}$. They are fixed by fitting the
properties of low energy mesons: $m_{\pi}=135.0\;\mathrm{MeV}$,
$m_{K}=497.7\;\mathrm{MeV}$, $m_{\eta^{\prime}}=957.8\;\mathrm{MeV}$
and $f_{\pi}=92.4\;\mathrm{MeV}$~\cite{Rehberg1996}.

We consider a homogeneous magnetic field $B$ along the z-direction.
In the mean field approximation, one can obtain the thermodynamical
potential density of the PNJL model in an external magnetic field.
Details of this calculation can be found in Ref.~\cite{Fu2011}. Here
we just give the final result as follows
\begin{eqnarray}
\Omega &=&-N_{c}\sum_{f=u,d,s}\frac{|q_{f}|e
B}{2\pi}\sum_{n=0}^{\infty}\alpha_{n}\int_{-\infty}^{\infty}\frac{d
p_{z}}{2\pi}\Bigg(f_{\Lambda}^{2}(p_{f})E_{f}+\frac{T}{3}\ln\Bigg\{1+3\Phi^{*}\exp\Big[-\Big(E_{f}\nonumber \\
&&-\mu_{f}\Big)/T\Big]+3\Phi
\exp\Big[-2\Big(E_{f}-\mu_{f}\Big)/T\Big]+\exp\Big[-3\Big(E_{f}-\mu_{f}\Big)/T\Big]\Bigg\}\nonumber\\
&&+\frac{T}{3}\ln\Bigg\{1+3\Phi
\exp\Big[-\Big(E_{f}+\mu_{f}\Big)/T\Big]+3\Phi^{*}\exp\Big[-2\Big(E_{f}+\mu_{f}\Big)/T\Big]\nonumber\\
&&+\exp\Big[-3\Big(E_{f}+\mu_{f}\Big)/T\Big]\Bigg\}\Bigg)+2G({\phi_{u}}^{2}
+{\phi_{d}}^{2}+{\phi_{s}}^{2})\nonumber\\
&&-4K\phi_{u}\,\phi_{d}\,\phi_{s}+\mathcal{U}(\Phi,\Phi^{*},T),\label{thermopotential}
\end{eqnarray}
where we have
\begin{equation}
p_{f}=\sqrt{2n|q_{f}|e B+p_{z}^{2}} \label{}
\end{equation}
and
\begin{equation}
E_{f}=\sqrt{2n|q_{f}|e B+p_{z}^{2}+M_{f}^{2}}.\label{Ef}
\end{equation}
The constituent mass is
\begin{equation}
M_{i}=m_{0i}-4G\phi_{i}+2K\phi_{j}\,\phi_{k},\label{constituentmass}
\end{equation}
where $\phi_{i}$ is the chiral condensate
$\langle\bar{\psi}\psi\rangle_{i}$. Since charged particles in the
lowest order Landau level are polarized by the external magnetic
field, the spin-degeneracy factor $\alpha_{n}$ in Eq.
(\ref{thermopotential}) is 1 for $n=0$ and 2 otherwise. Furthermore,
we should mention that when the magnetic field is strong, the sharp
cutoff usually used for the vacuum part in the PNJL model has a
problem to introduce cutoff artifact. To avoid this problem, we use
a smooth cutoff instead of the sharp one, which is realized by
introducing a cutoff function $f_{\Lambda}(p)$ in the vacuum part as
Eq. (\ref{thermopotential}) shows. We adopt the form of the cutoff
function in Ref.~\cite{Fukushima2010}, as given by
\begin{equation}
f_{\Lambda}(p)=\sqrt{\frac{\Lambda^{2N}}{\Lambda^{2N}+p^{2N}}}.
\label{}
\end{equation}
$N=10$ is chosen in our numerical calculations. As one can see,
$f_{\Lambda}(p)$ is in fact the sharp cutoff function
$\theta(\Lambda-p)$ in the limit $N\rightarrow\infty$.

From stationary conditions, we obtain a set of equations of motion
by Minimizing the thermodynamical potential in Eq.
(\ref{thermopotential}) with respect to three-flavor quark
condensates, Polyakov loop $\Phi$ and its conjugate $\Phi^{*}$.
These equations of motion can be solved as functions of temperature
$T$, strength of magnetic field $B$, three-flavor quark chemical
potentials $\mu_{u}$, $\mu_{d}$, and $\mu_{s}$ or conserved charge
chemical potentials $\mu_{B}$, $\mu_{Q}$, and $\mu_{S}$.

Substituting solutions of the equations of motion into Eq.
(\ref{thermopotential}), one can also obtain the thermodynamical
potential density and the pressure ($P=-\Omega$) of a
thermodynamical system in the mean field approximation. Then we can
calculate the derivatives of the pressure with respect to three
conserved charge chemical potentials, i.e.,
\begin{equation}
\chi_{ijk}^{BQS}=\frac{\partial^{i+j+k}(P/T^{4})}
{\partial(\mu_{B}/T)^{i}\partial(\mu_{Q}/T)^{j}\partial(\mu_{S}/T)^{k}},\label{susceptibility}
\end{equation}
which generalizes the quark number susceptibility to a more general
expression. In fact, the generalized susceptibilities $\chi$ 's in
Eq. (\ref{susceptibility}) are related to the cumulants of the
conserved charge multiplicity distributions, which can be observed
in heavy ion collision experiments. For example, the relations
between the second and higher order susceptibilities and the
fluctuations of conserved charges are given by
\begin{eqnarray}
\chi_{2}^{X}&=&\frac{1}{VT^{3}}\langle {\delta N_{X}}^{2} \rangle,  \\
\chi_{4}^{X}&=&\frac{1}{VT^{3}}\Big(\langle {\delta
N_{X}}^{4}\rangle-3{\langle {\delta N_{X}}^{2} \rangle}^{2}\Big), \\
\chi_{6}^{X}&=&\frac{1}{VT^{3}}\Big(\langle {\delta
N_{X}}^{6}\rangle-15\langle {\delta N_{X}}^{4}\rangle\langle {\delta
N_{X}}^{2}\rangle-10{\langle {\delta
N_{X}}^{3}\rangle}^{2}+30{\langle {\delta
N_{X}}^{2}\rangle}^{3}\Big),
\end{eqnarray}
where $\delta N_{X}=N_{X}-\langle N_{X}\rangle$ ($X=B,Q,S$) and
$\langle N_{X}\rangle$ is the ensemble average of the conserved
charge number $N_{X}$. V is the volume of the system. In the same
way, the mixed cumulants of conserved charge distributions, i.e.,
the correlations among conserved charges, can also be expressed as
their corresponding generalized susceptibilities, e.g.,
\begin{equation}
\chi_{11}^{XY}=\frac{1}{VT^{3}}\langle \delta N_{X}\delta
N_{Y}\rangle.
\end{equation}
In this work, we only consider the cases with $\mu_{B,Q,S}=0$, in
which the generalized susceptibilities are nonvanishing only when
$i+j+k$ is even.

\section{Chiral and deconfinement phase transitions}
\label{phasetransition}

\begin{figure}[!htb]
\includegraphics[scale=1.2]{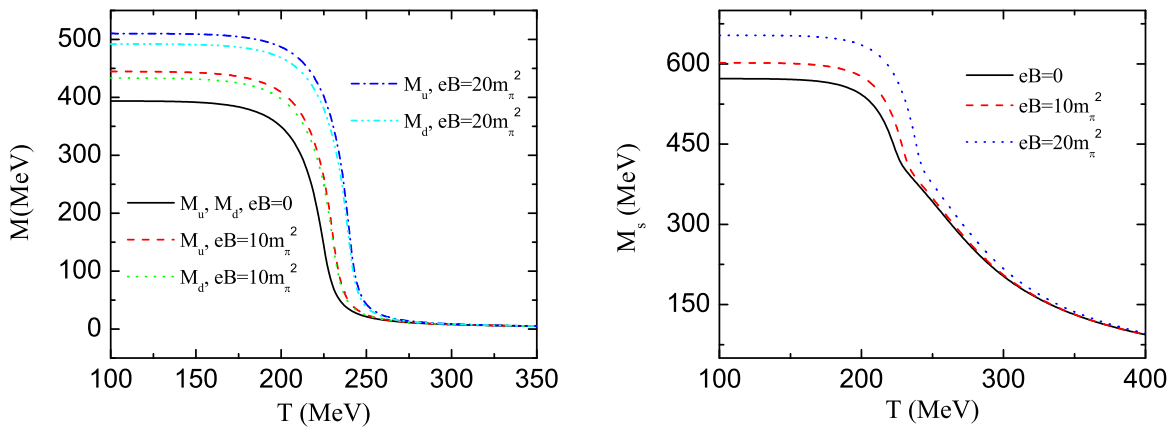}
\caption{(color online). Constituent masses of quarks as functions
of the temperature calculated in the PNJL model with several values
of $eB$ in unit of $m_{\pi}^{2}$. Left panel is for the $u$ and $d$
light quarks, and right panel is for the strange quarks.
$\mu_{B,Q,S}=0$ is chosen throughout our work.}\label{FigM}
\end{figure}

In this section we focus on the chiral and deconfinement phase
transitions of the 2+1 flavor PNJL model with an external magnetic
field. Figure~\ref{FigM} shows the chiral phase transition, where
the constituent quark masses are plotted as functions of the
temperature at several values of the magnetic field strength. We
find that the pseudocritical temperature for the chiral restoring
phase transition, which is defined by the position of the peak of
$|d M_{u,d}/dT|$ as a function of $T$, increases from 224 to
$240\,\mathrm{MeV}$, when $eB$ is increased from 0 to
$20m_{\pi}^{2}$. It is also found that the constituent quark masses
and the chiral condensates increase with $B$ at a given temperature,
which is more pronounced at low temperature as Fig.~\ref{FigM}
shows. The dependence of the chiral phase transition temperature on
the strength of an external magnetic field obtained in the 2+1
flavor PNJL model, are consistent with former effective model
computations of two flavor
systems~\cite{Fukushima2010,Mizher2010,Skokov2012}, and the expected
magnetic field-temperature phase diagram of QCD (as shown in Fig.1
in Ref.~\cite{Mizher2010}) where the chiral critical temperature
increases with $B$, while the deconfining one decreases with
increasing $B$. Like various effective models, our result is in
conflict with that of the state-of-the-art lattice
simulations~\cite{Bali2012}, where it was predicted that the chiral
critical temperature decreases with increasing magnetic field. The
left panel of Fig.~\ref{FigM} also shows an interesting result for
light quarks: there is a split between the $u$ and $d$ constituent
masses when the magnetic field is nonvanishing, and this split
becomes more prominent with the increase of $B$. The split between
light quarks at finite $B$ indicates that the $SU(2)$ symmetry
between $u$ and $d$ quarks is broken under the influence of an
external magnetic field, since the $u$ and $d$ quarks have different
electric charges. But this broken effect is not significant even
$eB$ is increased to the maximal value $20m_{\pi}^{2}$ of our
calculations, as Fig.~\ref{FigM} shows.

\begin{figure}[!htb]
\includegraphics[scale=0.8]{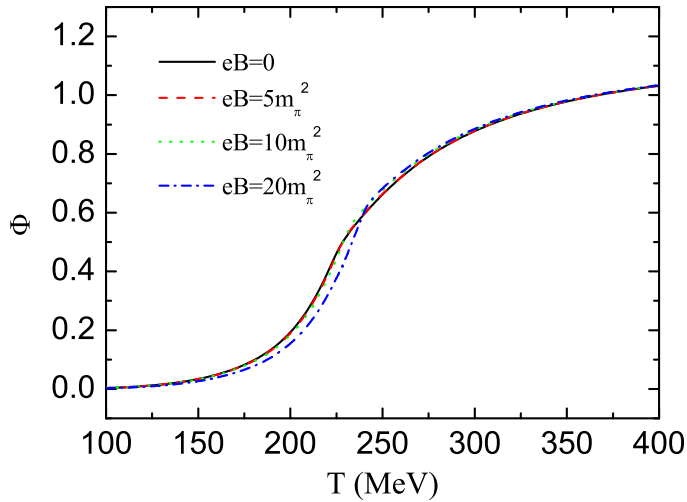}
\caption{(color online). Polyakov loop as a function of $T$ at
several values of $eB$.}\label{FigPhi}
\end{figure}

In Fig.~\ref{FigPhi} we show the deconfinement phase transition at
several values of $eB$, which is characterized by the dependence of
the deconfining order parameter, i.e., the Polyakov loop, on the
temperature. One can see that the impact of the magnetic field on
the Polyakov loop dynamics is much smaller than that on the chiral
phase transition, in particular, when the temperature is high, where
there is almost no difference among the several curves in
Fig.~\ref{FigPhi}. However, since the Polyakov loop dynamics is
entangled with the chiral one, especially at low temperature, the
pseudocritical temperature for the deconfinement phase transition
increases a little with $B$ as well, which is inconsistent with
lattice simulations~\cite{Bali2012} and the expected magnetic
field-temperature phase diagram of QCD. This problem also appears in
former effective model calculations~\cite{Fukushima2010,Mizher2010}.
This insufficiency of these effective models is due to the fact that
in these models, the Polyakov-loop effective potential which governs
the Polyakov-loop behavior is introduced by hand. The dynamical
couplings between the gluon field and the external magnetic field at
large $B$, are not included in these
models~\cite{Galilo2011,Fukushima2012}.

\begin{figure}[!htb]
\includegraphics[scale=1.2]{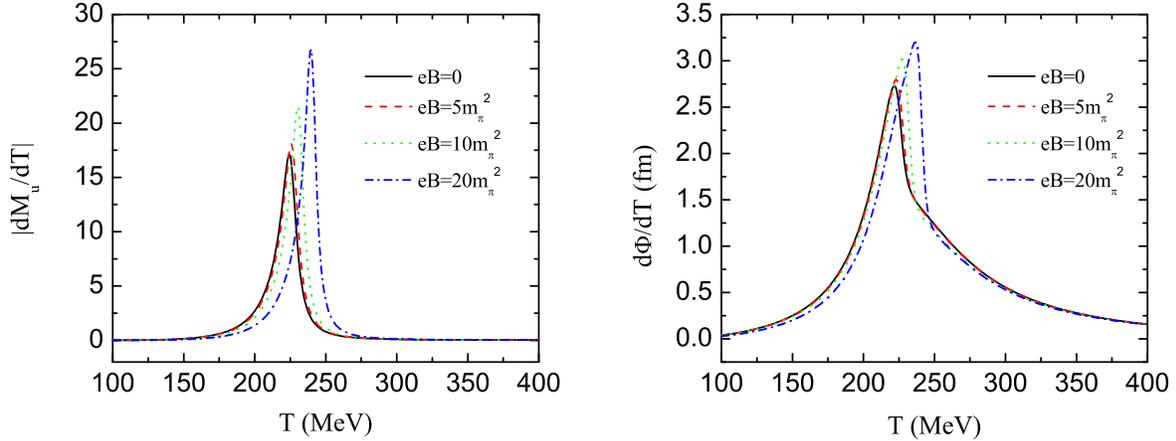}
\caption{(color online). $|d M_{u}/d T|$ (left panel) and $d \Phi/ d
T$ (right panel) versus temperature at several values of
$eB$.}\label{FigdMdT}
\end{figure}

Figure~\ref{FigdMdT} shows the derivatives of the $u$ quark
constituent mass and the Polyakov loop with respect to $T$ at
several values of $eB$. We observe that the peak value of $|d
M_{u}/d T|$ increases with $B$, which indicates that the chiral
phase transition becomes sharper when $B$ is increased. In another
word, the strength of the chiral phase transition increases with
$B$. This result agrees with two flavor effective model
calculations~\cite{Mizher2010,Skokov2012} and lattice
simulations~\cite{Bali2012}. Furthermore, we also find that the peak
value of $d \Phi/ d T$ increases a little with $B$.

To summarize the results in this section, we should mention that
effective models and the state-of-the-art lattice simulations give
different results on how the chiral and deconfining critical
temperatures are influenced by an external magnetic field, and the
interplay between the external magnetic field and the gluonic
dynamics and that between the magnetic field and the chiral dynamics
are not fully understood. However, effective models and lattice
simulations all consistently predict that the strength of the phase
transition increases with $B$. We mainly focus on fluctuations and
correlations of quarks and conserved charges in an external magnetic
field in this work. We will show below that our conclusions are
mainly based on the fact that the transition strength of the
crossover increases with $B$. Therefore, we expect that results
about the fluctuations and correlations obtained in the 2+1 flavor
PNJL model are reliable.

\section{Fluctuations of quarks}
\label{fluctuationsquark}

\begin{figure}[!htb]
\includegraphics[scale=0.60]{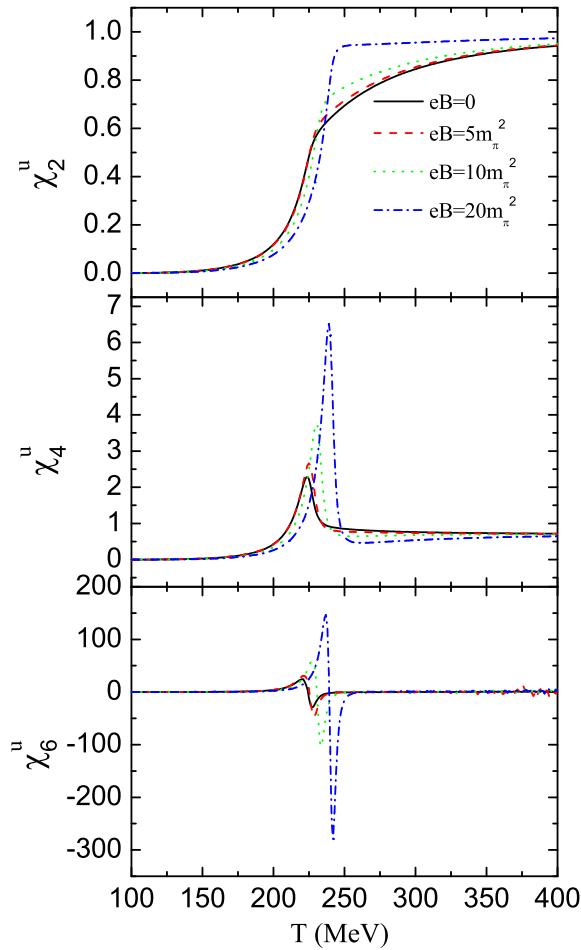}
\caption{(color online). Quadratic (top), quartic (middle), and
sixth-order (bottom) fluctuations of $u$ quarks as functions of the
temperature calculated in the PNJL model with several values of
$eB$.}\label{Figchiu}
\end{figure}

In this work we employ the method of Taylor expansion to calculate
the fluctuations and correlations given in Eq.
(\ref{susceptibility}). First of all, we focus on the influences of
an external magnetic field on the fluctuations of quarks. We plot
quadratic, quartic and sixth-order fluctuations of $u$ quarks as
functions of $T$ at several values of $eB$ in Fig.~\ref{Figchiu}.
One can see that the quadratic fluctuations $\chi^{u}_{2}$ above the
pseudocritical temperature increase with $B$. Furthermore, the
evolution of $\chi^{u}_{2}$ with $T$ during the chiral crossover
becomes sharper as $B$ is increased. Similar results are also found
in the computations of $\chi^{u}_{4}$ and $\chi^{u}_{6}$. We find
that the peak value of quartic fluctuations and the amplitude of
oscillations of $\chi^{u}_{6}$ during the chiral crossover are
greatly enhanced with the increase of $B$. As we have found in the
last section that the transition strength of the crossover increases
with $B$, it is natural to expected that the fluctuations increase
with $B$ as well, since fluctuations are closely related with the
strength of a crossover. When the crossover becomes an exact
second-order phase transition, the fluctuations should be divergent.

\begin{figure}[!htb]
\includegraphics[scale=1.2]{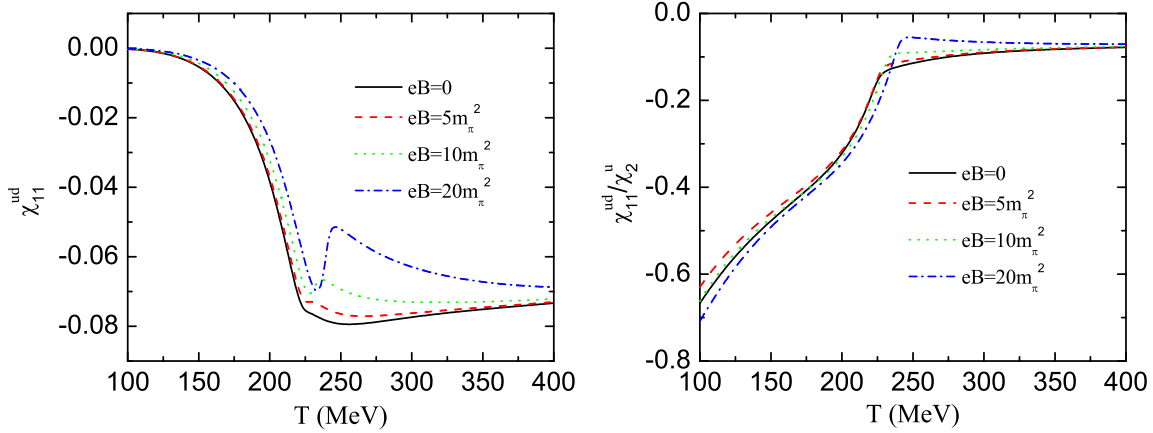}
\caption{(color online). Second-order correlation between $u$ and
$d$ quark numbers $\chi_{11}^{ud}$ (left panel) and its ratio with
respect to $\chi_{2}^{u}$ (right panel) as functions of $T$ at
several values of $eB$.}\label{Figchi1u1d}
\end{figure}

We show the mixed susceptibility between $u$ and $d$ quark numbers,
i.e., the second-order correlation between light quarks, at several
values of $eB$ in Fig.~\ref{Figchi1u1d}. Our calculated result
indicates that the value of $\chi_{11}^{ud}$ is negative, and there
is an inflection point at the pseudocritical temperature in the
curve of $\chi_{11}^{ud}$ as a function of $T$ when there is no
magnetic field. With the increase of $B$, the inflection point
develops a complicated dependent behavior on $T$, where
$\chi_{11}^{ud}$ oscillates during the crossover. In order to
compare $\chi_{11}^{ud}$ with $\chi_{2}^{u}$, we plot their ratio in
the right panel of Fig.~\ref{Figchi1u1d}. One can observe that
$\chi_{11}^{ud}$ is comparable to $\chi_{2}^{u}$ when the
temperature is below the pseudocritical temperature. However, when
the temperature is high, $\chi_{11}^{ud}$ can be neglected compared
with $\chi_{2}^{u}$, since $\chi_{11}^{ud}\rightarrow 0$ in the
Stefan-Boltzmann limit at high temperature.

\begin{figure}[!htb]
\includegraphics[scale=1.2]{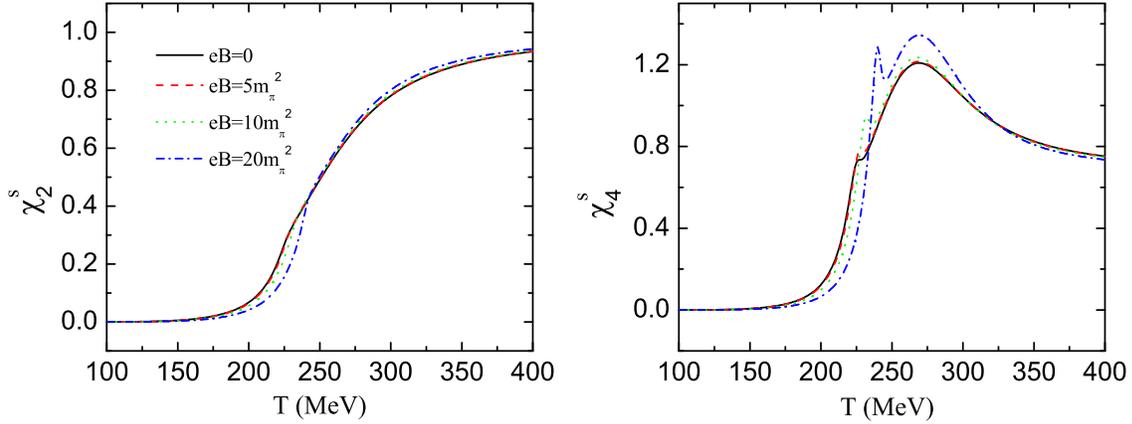}
\caption{(color online). Quadratic (left panel) and quartic (right
panel) fluctuations of $s$ quarks as functions of the temperature at
several values of $eB$.}\label{Figchis}
\end{figure}

\begin{figure}[!htb]
\includegraphics[scale=1.2]{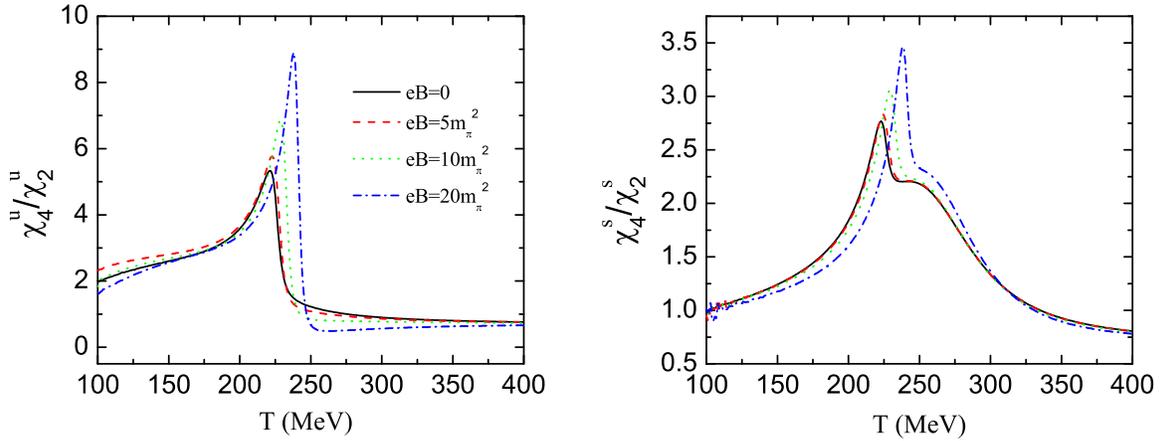}
\caption{(color online). Ratio of the quartic to quadratic
fluctuations for $u$ quarks (left panel) and $s$ quarks (right
panel) as functions of $T$ at several values of
$eB$.}\label{Figratiochius}
\end{figure}

Figure~\ref{Figchis} shows the evolution of quadratic and quartic
fluctuations of $s$ quarks with $T$ during the chiral crossover. In
the same way, we choose several values of $eB$. The influence of the
magnetic field on the fluctuations of $s$ quarks are much smaller
than that on light quarks, in particular for the low order
fluctuations $\chi_{2}$. One can see that there is almost no
difference among the several curves corresponding to different
values of $eB$ in the left panel of Fig.~\ref{Figchis}. But with the
increase of the order of fluctuations, the impact of magnetic field
becomes more pronounced as shown in the right panel of
Fig.~\ref{Figchis}. One observes that there are two peaks on the
curves of $\chi_{4}^{s}$ as a function of $T$, which correspond to
chiral restorations for light quarks and strange quarks,
respectively. More detailed discussions about this can be found in
Ref.~\cite{Fu2008}.

It is usually believed that the ratio of the quartic to quadratic
fluctuations of quarks is a valuable probe of the deconfinement and
chiral phase
transitions~\cite{Ejiri2006a,Ejiri2006b,Karsch2006,Stokic2009}.
Because there is a pronounced peak in the curve of the ratio at the
critical temperature. In Fig.~\ref{Figratiochius} we show this ratio
for $u$ and $s$ quarks with several values of $eB$. One observes
that both the peak in $\chi_{4}^{u}/\chi_{2}^{u}$ and that in
$\chi_{4}^{s}/\chi_{2}^{s}$ increase with the strength of magnetic
field. We should emphasize that since the ratio deducts the
influence of the phase space change resulting from the Landau levels
in an external magnetic field, the increase of the ratio is not due
to the phase space, but to the increase of the transition strength
of the crossover. More discussions about the ratio
$\chi_{4}/\chi_{2}$ can be found in Refs.~\cite{Skokov2010,Fu2010a}.
It should be mentioned that the ratio in Ref.~\cite{Skokov2010} is
not $\chi_{4}^{u}/\chi_{2}^{u}$ calculated here, but is more closely
related to $\chi_{4}^{B}/\chi_{2}^{B}$ which will be discussed in
the following (there is a difference of factor 9).

\section{Fluctuations and correlations of conserved charges}
\label{conservedcharges}

In this section we discuss fluctuations and correlations of
conserved charges, e.g., baryon number, electric charge, and
strangeness. These charges are conserved throughout the evolution of
a fire ball produced in relativistic heavy ion collisions.
Therefore, it is expected that fluctuations of these conserved
charges and correlations among them, which can be extracted from
event-by-event fluctuations in
experiments~\cite{Jeon2004,Koch2005,Abelev2009}, carry information
about the properties of the fire ball during its early evolution
stage, including the strong and quickly decaying magnetic field
produced in noncentral collisions.

\begin{figure}[!htb]
\includegraphics[scale=0.60]{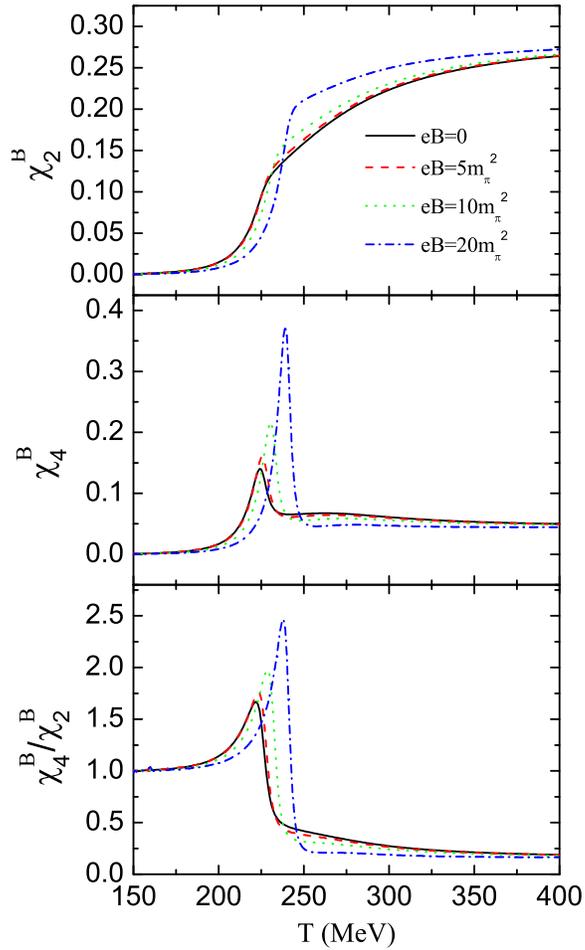}
\caption{(color online). Quadratic (top) and quartic (middle)
fluctuations of baryon number, and their ratio (bottom) as functions
of $T$ at several values of $eB$.}\label{FigchiB}
\end{figure}

\begin{figure}[!htb]
\includegraphics[scale=0.60]{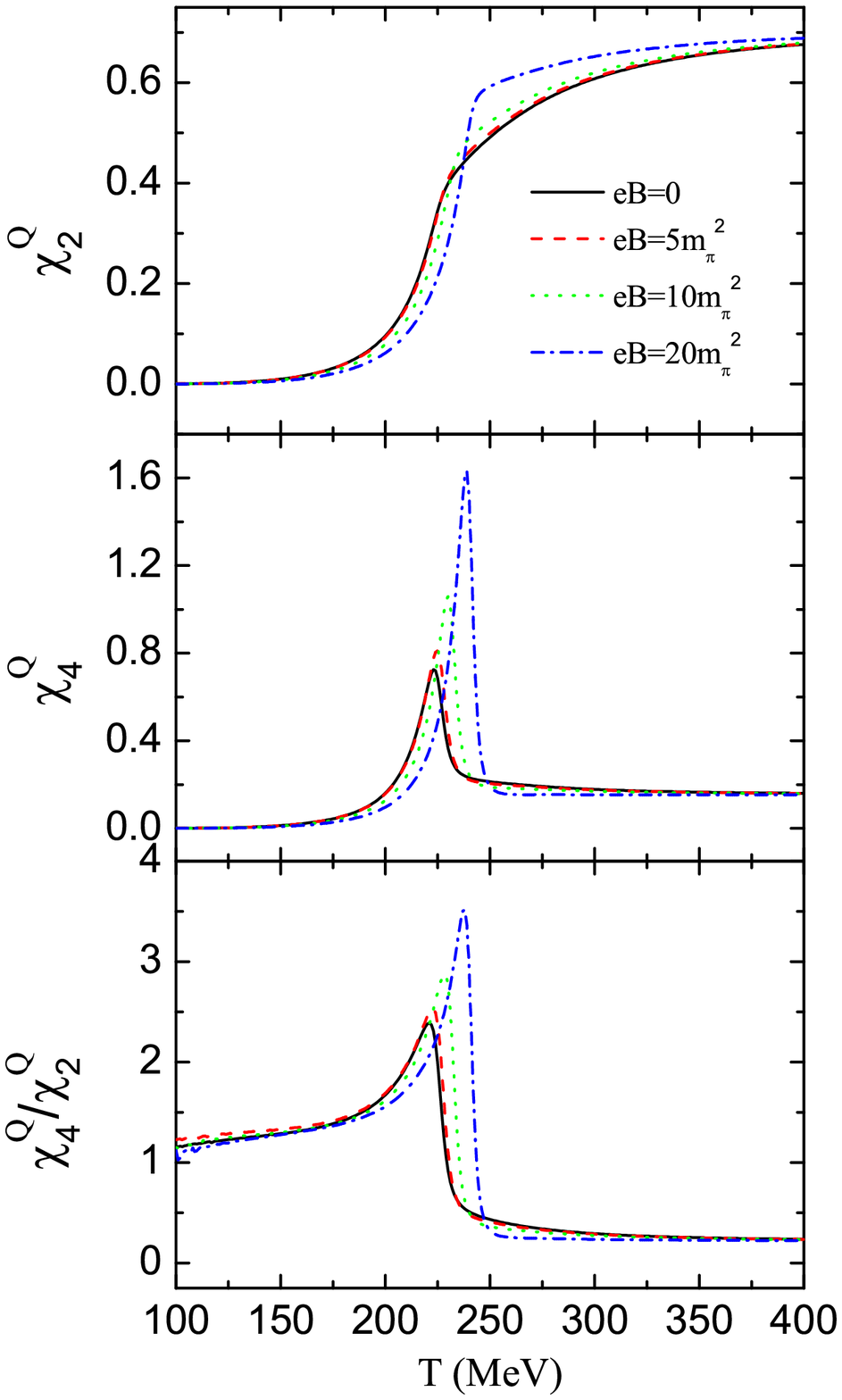}
\caption{(color online). Quadratic (top) and quartic (middle)
fluctuations of electric charge, and their ratio (bottom) as
functions of $T$ at several values of $eB$.}\label{FigchiQ}
\end{figure}

Figure~\ref{FigchiB} and figure~\ref{FigchiQ} shows fluctuations of
baryon number and electric charge, respectively. The quadratic and
quartic fluctuations and their ratios are plotted as functions of
$T$ at several values of $eB$. Like the quadratic fluctuations of
quarks, we find that the transition of $\chi_{2}^{B}$
($\chi_{2}^{Q}$) with $T$ during the chiral crossover becomes
sharper when the strength of magnetic field is increased. The
quartic fluctuations of baryon number and electric charge  present a
pronounced peak at the pseudocritical temperature, and the peak
value increases with increasing $B$. In the same way, we also show
$\chi_{4}^{B}/\chi_{2}^{B}$ and $\chi_{4}^{Q}/\chi_{2}^{Q}$ versus
temperature. Both the peak values of these two ratios increase with
the magnetic field strength, which demonstrates that since the
transition strength of the crossover increases with $B$, the
fluctuations of conserved charges during the chiral crossover are
enhanced when the strength of magnetic field is increased. Comparing
Fig.~\ref{FigchiQ} with Fig.~\ref{FigchiB}, we find that the
magnitude of electric charge fluctuations is larger than that of
baryon number fluctuations.

\begin{figure}[!htb]
\includegraphics[scale=0.60]{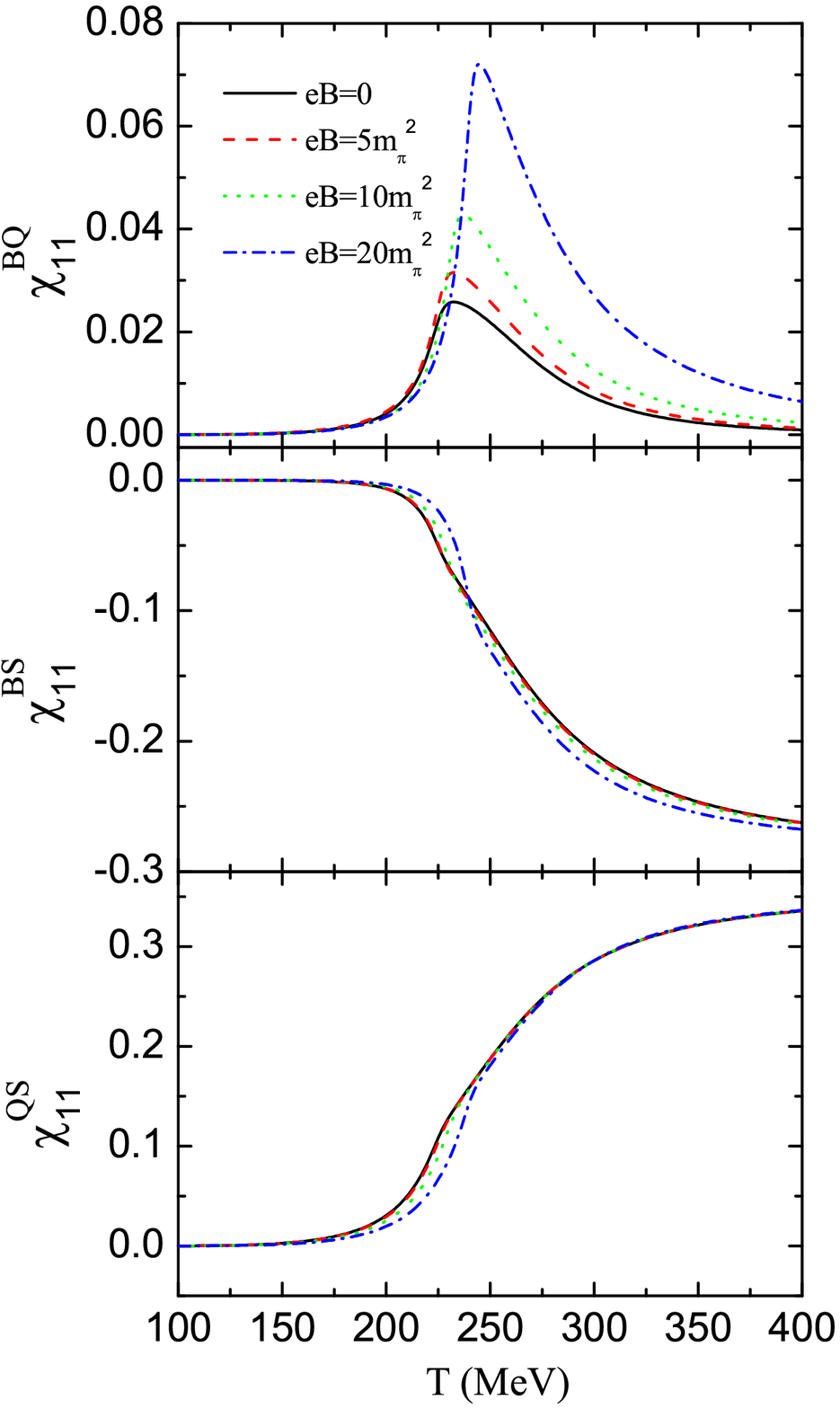}
\caption{(color online). Correlations between baryon number and
electric charge (top), baryon number and strangeness (middle),
electric charge and strangeness (bottom) as functions of the
temperature with several values of $eB$.}\label{Figchi1X1Y}
\end{figure}

In Fig.~\ref{Figchi1X1Y} we show the correlations $\chi_{11}^{BQ}$,
$\chi_{11}^{BS}$, and $\chi_{11}^{QS}$ as functions of $T$ at
several values of $eB$. We find that the impact of the external
magnetic field on $\chi_{11}^{BS}$ and $\chi_{11}^{QS}$ is small,
like quadratic fluctuations of conserved charges discussed above.
However the correlations between baryon number and electric charge
are much more sensitive to the magnetic field. One can observe that
there is a bump on the curve of  $\chi_{11}^{BQ}$ versus
temperature. The bump becomes sharper with increasing $B$, and the
height of the bump increases rapidly with $B$. Since the mass of
strange quarks is much larger than that of light quarks, the
response of strange quark fluctuations and correlations to the
external magnetic field is less sensitive than that of light quarks,
which is also presented in the calculations of $\chi_{2}^{s}$ and
$\chi_{4}^{s}$ in Fig.~\ref{Figchis}. Therefore, the dependence of
$\chi_{11}^{BQ}$ on $B$ is more pronounced than those of
$\chi_{11}^{BS}$ and $\chi_{11}^{QS}$. Another reason is that
$\chi_{11}^{BQ}$ is vanishing in the Stefan-Boltzmann limit at high
temperature, while $\chi_{11}^{BS}$ and $\chi_{11}^{QS}$ have finite
values in this limit. So the dependence of $\chi_{11}^{BS}$ and
$\chi_{11}^{QS}$ on $B$ may be polluted by the background, while
$\chi_{11}^{BQ}$ has no such problem.

\section{Conclusions}
\label{summary}

In this work, we have studied influences of an external magnetic
field on the deconfinement and chiral restoring phase transitions in
the 2+1 flavor PNJL model. Calculations of the 2+1 flavor PNJL model
indicate that magnetic field catalyze the dynamical chiral symmetry
breaking, and the chiral pseudocritical temperature increases with
increasing $B$. The deconfinement pseudocritical temperature
increases a little with $B$ as well. We find that the transition
strength increases when $B$ is increased.

Effects of the external magnetic field on the fluctuations of quark
number are studied in the effective model. We find that the magnetic
field makes the transition of the quadratic fluctuations with
respect to $T$ during the chiral crossover sharper, since the
transition strength of the crossover increases with $B$. The peak
structure in quartic fluctuations and the oscillation in sixth-order
fluctuations become more and more prominent when the magnetic field
strength is increased.  With the increase of $B$, the inflection
point in the curve of $\chi_{11}^{ud}$ develops a complicated
dependent behavior on $T$, where $\chi_{11}^{ud}$ oscillates during
the crossover. Comparing the fluctuations of strange quarks with
those of light quarks, we find the influences of an external
magnetic field on strange quark fluctuations are smaller than those
on light quarks.

Special attentions are paid on the fluctuations and correlations of
conserved charges, including baryon number, electric charge and
strangeness, Since they can be measured with event-by-event
fluctuations in heavy ion collision experiments. In the same way, we
find that the transition of quadratic fluctuations of conserved
charges becomes more abrupt when $B$ is increased. The peak value of
quartic fluctuations of conserved charges increases with $B$. We
should emphasize that the peak structure in the ratios
$\chi_{4}^{B}/\chi_{2}^{B}$ and $\chi_{4}^{Q}/\chi_{2}^{Q}$ become
more and more pronounced with the increase of $B$, which indicates
that $\chi_{4}^{B}/\chi_{2}^{B}$ and $\chi_{4}^{Q}/\chi_{2}^{Q}$ may
be useful probes for the strong magnetic field produced in early
noncentral collisions. We also study the correlations of conserved
charges.  We find that the impact of the external magnetic field on
$\chi_{11}^{BS}$ and $\chi_{11}^{QS}$ is small, but $\chi_{11}^{BQ}$
are sensitive to the magnetic field.

\section*{Acknowledgements}
This work was supported by the National Natural Science Foundation
of China under Contracts No. 11005138.


\end{document}